\documentstyle[titlepage,12pt]{article}

\def\be{\begin{equation}}
\def\bea{\begin{eqnarray}}
\def\ee{\end{equation}}
\def\eea{\end{eqnarray}}
\def\sgn{\mbox{sgn}}

\newcommand{\half}      {\frac{1}{2}}
\newcommand{\R}{\mbox{$I\!\!R$}}

\newcommand{\pda}[2]    {\frac{\partial   #1}{\partial #2  }}
\newcommand{\pdb}[2]    {\frac{\partial^2 #1}{\partial #2^2}}
\newcommand{\pdc}[3]    {\frac{\partial^2 #1}{\partial #2 \partial #3}}

\newcommand{\tda}[2]    {\frac{d   #1}{d #2  }}
\newcommand{\tdb}[2]    {\frac{d^2 #1}{d #2^2}}

\newcommand{\st}        {\:|\:}

\newcommand{\req}[1]    {(\ref{#1})}

\newcommand{\ie}        {{\em i.e.}}

\newcount\sectionnumber
\def\sect
{\global\equationnumber=0
\global\advance\sectionnumber by 1
\the\sectionnumber . }

\newcount\equationnumber
\def   \num
{\eqno{\global\advance\equationnumber by 1
\left(\the\sectionnumber .\the\equationnumber \right)}}

\newtheorem{thm}{Theorem}

\newenvironment{proof}{\noindent {\bf Proof:}}{\hfill $\Box$}

\begin{document}

\title{The Causal Structure of Two-Dimensional Spacetimes}

\author{Dan Christensen and Robert B. Mann\\
	Department of Physics\\
	University of Waterloo\\
	Waterloo, Ontario\\
	Canada N2L 3G1}

\date{October 31, 1991\\WATPHYS TH-91/05\\to appear in Class. Quant. Grav.}

\maketitle

\begin{abstract}
  We investigate the causal structure of $(1+1)$-dimensional spacetimes.
  For two sets of field equations we show that at least locally
  any spacetime is a solution for an appropriate choice of the
  matter fields. For the theories under consideration we
  investigate
  how smoothness of their black hole solutions affects
  time orientation.  We show that if an analog to Hawking's area theorem
  holds in two spacetime dimensions, it must actually state
  that the size of a black
  hole never {\em increases}, contrary to what happens in four dimensions.
  Finally, we discuss the applicability of the Penrose and
  Hawking singularity theorems to two spacetime dimensions.
\end{abstract}

\section{Introduction}

Relativistic theories of gravitation in two spacetime dimensions provide an
interesting theoretical laboratory for understanding issues relevant to
quantum gravity.  Such theories reduce the complexity
of $(3+1)$-dimensional general relativity significantly, thereby offering
much hope for obtaining significant insights into its quantization, as well
as an understanding of the issues associated with short-distance problems,
topology change, singularities and the  cosmological constant problem.
Recent work has revealed interesting relationships between $(1+1)$-dimensional
gravitational theories  and conformal field theory \cite{Pol87}, the
Liouville  model \cite{Tei84,Jac84,Jac85}, random lattice models
\cite{Kni88}, and sigma models \cite{Leb88,Geg88b,Geg89,Lin87}.

Although Einstein's field equations are trivial in two spacetime dimensions,
there exist a variety of $(1+1)$-dimensional generally covariant
theories of gravitation
\cite{Tei84,Jac84,Jac85,Tei83,Hen85,Fuk85,Geg88a}
some of which have non-trivial dynamical structure.  A
close analog of the Einstein equations is given by  \cite{Man89,Man90}
  \begin{equation}
      R - \Lambda = 8 \pi G T
      \label{R=T}
  \end{equation}
along with the conservation equation
  \begin{equation}
      T^{\mu\nu}_{\ \;\;;\nu} = 0
      \label{cons}
  \end{equation}
where $R$ is the Ricci scalar, $\Lambda$ is a cosmological constant, $T$ is
the trace of the stress-energy tensor, $G$ is Newton's constant and we have
taken the speed of light to be unity.   These equations can be derived from
a local action principle \cite{Semi} by incorporating an auxiliary field
whose classical evolution does not affect the gravity/matter system above.
In the absence of matter this system reduces to the vacuum field equation
used in the Liouville model \cite{Tei84,Jac84,Jac85}.
The classical aspects of this theory of gravity have been examined in some
detail \cite{Man89,Man90,Sik91} and it has been shown that it has a
remarkable similarity to to four-dimensional general relativity in many of
its features. These features include a Newtonian limit, Robertson-Walker
cosmological solutions, interior solutions, gravitational waves and the
gravitational collapse of dust into a black hole with an event horizon
structure which is the same as that of the four-dimensional Schwarzschild
solution. Indeed, the field equations of this theory follow from a
dimensional reduction of Einstein's equations in a certain limit
\cite{Lowblack}; in this sense they form a $(1+1)$-dimensional version of
general relativity.

These classical features are so closely analogous to $(3+1)$-dimensional
general relativity that one might hope its quantization would bear a
similar resemblance to $(3+1)$-dimensional quantum gravity. The
semiclassical properties of this theory \cite{Semi,SharDir,TomRobb} do
indeed yield interesting effects such as Hawking radiation and black hole
condensation.  These properties are intimately connected with the
non-trivial event horizon structures which can form in the theory in a
manner quite similar to their $(3+1)$-dimensional general relativistic
counterparts.

Much more recently it has been shown that other $(1+1)$-dimensional
theories of gravity which arise in the context of non-critical
string theory can also yield a non-trivial event horizon structure.
A spacetime exhibiting such features was recently
discovered as a solution to a scale-invariant
higher-derivative theory of gravity \cite{Sch91}, and was later
found to be a solution to $c=1$ Liouville gravity \cite{Lug91} as well
as to a non-critical string theory in two spacetime dimensions
\cite{MSW91,Wit91}. This latter result suggests the
possiblity of using string-theoretic technology to examine the
formation and properties of black holes in more realistic cases.

The field equations associated with this theory are
  \begin{equation}
    e^{-2 \phi} (R_{ab} + 2 \nabla_a \nabla_b \phi) = 8 \pi G T_{ab}
    \label{string1}
  \end{equation}
  \begin{equation}
    R - 4 (\nabla \phi)^2 + 4 \nabla^2 \phi + J + c = 0.
    \label{string2}
  \end{equation}
where a stress-energy tensor $T_{ab}$ and source $J$ for the dilaton
field $\phi$ have been included. For $J=0=T_{ab}$,
these equations reduce to those of non-critical $(1+1)$-dimensional string
theory in the absence of a tachyon field \cite{MSW91}.
The black hole metric which follows from (\ref{string1},\ref{string2})
in this case is unique; it is asymptotically flat, and
may be matched to a solution for collapsing dust provided appropriate
surface stresses are included, where  the source $J$ may be
understood to arise from the tachyon sector \cite{Ros91}.
The quantum properties of the above metric are similar to those
found in ref. \cite{Sik91}.

In this paper, we investigate the causal structure of solutions to
the above theories.  We begin by
motivating the various solutions in section~\ref{background}.
In section~\ref{restrict} we ask whether an
arbitrary spacetime can be considered to be a solution to the field equations
for an appropriate choice of the matter fields, and we
provide a partial answer.  In section~\ref{bh} we examine some
black hole solutions and look at their causal structure in detail.
We investigate how smoothness of solutions affects their time orientation.
We mention that if an analog to Hawking's area
theorem \cite{Haw73} holds in two spacetime dimensions, it must actually state
that the size of a black hole never {\em increases}, which is exactly
the opposite of what happens in four dimensions,
and we note the difficulty in interpreting the ``size'' of a black hole.
In section~\ref{sing} we discuss the applicability of the Penrose and
Hawking singularity theorems to two spacetime dimensions, focusing
on Penrose's 1965 theorem \cite{Haw73,Pen65} which predicts that a
spacetime containing a closed trapped surface must be singular.  We
explain the difficulty in defining a closed trapped surface in two
dimensions and show that the energy condition in Penrose's theorem is
trivially true for all $(1+1)$-dimensional spacetimes.  A spacetime with a
black hole which contains singularities of an
unexpected type is used as an illustrative example.
Finally, we summarize our results in a concluding section.

\section{Background and Motivation}
\label{background}

Throughout this paper we will be examining many different spacetimes, so
here we introduce each of them and describe the context in which they
arise.  In general our conventions follow those of Hawking and Ellis
\cite{Haw73}.

By a {\em spacetime} $(M,g_{ab})$ we shall mean a Hausdorff
$C^\infty$ manifold $M$ (without boundary) of dimension $\geq 2$ with a
non-degenerate Lorentzian metric $g_{ab}$,
that is, a metric of signature $(-,+,\cdots,+)$.
We define a vector $v^a$ to be {\em timelike}, {\em null} or
{\em spacelike} if $g_{ab} v^a v^b$ is negative, zero or positive,
respectively.  We also assume that the spacetime is time-orientable,
\ie\ that there exists a continuous timelike vector field on $M$.

We define the Riemann tensor, Ricci tensor and Ricci scalar as
  \begin{equation}
    R^a_{\;bcd} = \pda{\Gamma^a_{\;bd}}{x^c} - \pda{\Gamma^a_{\;bc}}{x^d}
      + \Gamma^a_{\;cf} \Gamma^f_{\;bd} - \Gamma^a_{\;df} \Gamma^f_{\;bc}
    \label{riemann}
  \end{equation}
  \begin{equation}
    R_{bd} = R^a_{\;bad}
    \label{ricci}
  \end{equation}
and
  \begin{equation}
    R = R^a_{\;a} = R^a_{\;bad} g^{bd}
    \label{rs}
  \end{equation}
respectively.

For our purposes it will be useful to write the static metric in the form
  \begin{equation}
    ds^2 = - \alpha(x) dt^2 + \frac{dx^2}{\alpha(x)}.
    \label{alpha}
  \end{equation}
Such a choice of coordinates is always possible, at least locally.
Spacetimes given by static metrics of this form fall into four
distinct categories distinguished by the sign of $\alpha$ at large $|x|$:
\be
\lim_{|x|\to\infty} \sgn(\alpha(x)) =  \left\{ \begin{array}{ll}
		    +1 & \mbox{case (A)}  \\
		    -1 & \mbox{case (B)}  \\
		    \sgn(x) & \mbox{case (C)}  \\
		    \mbox{no limit} & \mbox{case (D)}
\end{array} \right. .   \label{eq13}
\ee
Case (A) is the spacetime one would expect to arise from the endpoint
of gravitational collapse of a distribution of $(1+1)$-dimensional matter.
Before collapse the signature of the metric is everywhere $(-,+)$, but
afterward certain regions of spacetime develop event horizons.
In this case $\alpha$ must have an even number of roots (some pairs of
which may be degenerate)
We shall cite an example of this below.
Physical $(1+1)$-dimensional observers (\ie\ those abiding in
a spacetime of signature $(-,+)$, where $t$ is timelike) may be located at
regions of large $x$, but will ultimately be unable to receive signals from
observers at large $-x$ since all such signals must cross the event horizon
once collapse has occurred. The second case is what one might expect in a
spacetime which had a cosmological constant, as we will illustrate below.
Again, $\alpha$ must have an even number of roots. In this situation,
observers are located only in regions of $|x|< R_0$ (where $R_0$ is some
constant),  and are unable to receive information from more distant regions
of their universe. Case (C) is somewhat unusual in that it has no
$(3+1)$-dimensional analogue:
spacetime has signature $(-,+)$ for large $x$ and has
signature $(+,-)$ for large $-x$;   without loss of generality $x$ may be
taken to be positive as above.
Originally such spacetimes were considered
in the context of a  higher-derivative theory \cite{Sch91}; more recently
they have become of interest in the context of finding solutions to the
system (\ref{string1},\ref{string2}) with $J=T_{ab}=0$ \cite{MSW91,Wit91}.
Finally, case (D) involves those spacetimes for which $\alpha$ has no definite
sign for large $|x|$.

We begin with solutions to the system \req{R=T} and \req{cons}.
The symmetric, continuous and static solutions for a point particle
situated at the origin are given by \req{alpha} where
  \begin{equation}
    \alpha(x) = - \half \Lambda x^2 + 2 M |x| - C
    \label{point-source}
  \end{equation}
on $\R^2$.
$M$ can be interpreted as the mass of the source and
$C$ is an arbitrary (but meaningful) constant.
At $x = 0$ the metric is continuous but not differentiable for $M \ne 0$.
When $\alpha(x) = 0$ the metric is singular, but for $x \ne 0$
these are just coordinate singularities that result from writing
the metric in the form \req{alpha}.  For various choices of $\Lambda$,
$M$ and $C$ the spacetime represents a black hole, a white hole,
a naked singularity, or other more complicated structures.  This
spacetime can also be easily extended to multiple point sources.
All of this is explored in detail in \cite{Man90}.

Spacetimes with $\Lambda\neq 0$ can have a qualitatively different
structure than their $(3+1)$-dimensional counterparts.
In $(3+1)$ dimensions the most general static isotropic metric may be
written in  the form
\begin{equation}
ds^2 = -B(r) dt^2 + A(r) dr^2 +  r^2 \left( d\theta^2
+ \sin^2\theta d\phi^2 \right).
\end{equation}
The $rr$ and $tt$ equations of general relativity imply that $AB$ is a
positive constant $C$ and that
\begin{equation}
-\Lambda = \half \tdb{\tilde{B}}{r} +
\frac{1}{r} \tda{\tilde{B}}{r} \label{r-r},
\end{equation}
where $\tilde{B} = B/C$, whereas the $\theta\theta$ and $\phi\phi$ parts of
Einstein's equations imply
\begin{equation}
-r^2 \Lambda = -1 + r \tda{\tilde{B}}{r} + \tilde{B} \quad . \label{th-th}
\end{equation}
As \req{th-th} implies \req{r-r}, we obtain the solution
\begin{equation}
\tilde{B}(r) = \left( 1-\frac{\Lambda}{3}r^2 \right) + \frac{c_3}{r}
\label{3+1Lam}
\end{equation}
which is the $(3+1)$-dimensional analog of \req{point-source}.
The metric \req{3+1Lam} is of type (B) above.
Note that the ratio between the $r^2$ coefficient and the constant term is
forced by \req{th-th} to be $-\Lambda/3$, in contrast to the freedom
available in choosing $C$ in the $(1+1)$-dimensional solution
\req{point-source}.
This freedom arises because of the lack of angular information in
$(1+1)$ dimensions; there is no equation corresponding to \req{th-th}.
Consequently a cosmological event horizon can only arise for $\Lambda>0$ in
$(3+1)$ dimensions, whereas such horizons can appear for either sign of
$\Lambda$ in \req{point-source}. Hence in $(1+1)$ dimensions, metrics
with cosmological event horizons of type (A) and (B) are both possible.

Another interesting solution to equations \req{R=T} and \req{cons}
arises in the symmetric collapse of an initially static
distribution of pressureless dust \cite{Sik91}.
The metric for the interior of the dust is given by
  \begin{equation}
    ds^2 = - dt^2 + (1-b t^2)^2 dx^2
    \label{R=T-dust}
  \end{equation}
in the region $S = \{(t,x) \st 0 \leq t < 1/\sqrt{b}, |x| <= r\}$,
where $x = \pm r$
are the (constant) positions of the edges of the dust in the comoving
coordinate system and $b \equiv 2 \pi G \rho_0$.
The dust collapses from a static condition of uniform density
$\rho_0$ at $t = 0$
to one of infinite density as $t \rightarrow 1/\sqrt{b}$.  To match
this metric to an external vacuum solution we define
  \begin{equation}
    X(t,x) = x(1-b t^2)
    \label{trans1}
  \end{equation}
  \begin{equation}
    T(t,x) = \frac{1}{2br} \tanh^{-1} \left[
      \frac{2brt}{\sqrt{bt^2+e^{2b(x^2-r^2)}(1-bt^2)}} \right].
    \label{trans2}
  \end{equation}
The above transformation is a one-to-one transformation of $S$
if $br^2 \leq 1/4$; otherwise there are cases in which $(t_1,x)$ and
$(t_2,x)$ map to the same $(T,X)$. However if we restrict ourselves to
$t < 1/2br$ the transformation will again be one-to-one.
Note that this transformation always transforms the boundary $x = \pm r$
in a one-to-one manner.
In $(T,X)$ coordinates the metric can be written
  \begin{equation}
    ds^2 = -B(T,X) dT^2 + \frac{dX^2}{1-4 b^2 x^2 t^2}
  \end{equation}
where
  \begin{equation}
 B(T,X) = \frac{\left[ bt^2 + e^{2b(x^2-r^2)}(1-bt^2)-4b^2 r^2 t^2 \right]^2
      \left[ bt^2 + e^{2b(x^2-r^2)}(1-bt^2) \right]}
      {e^{4b(x^2-r^2)} \left[ 1-4b^2 x^2 t^2 \right]}
  \end{equation}
and in which $x$ and $t$ are defined implicitly by \req{trans1} and
\req{trans2}.  This matches the static flat outside metric
  \begin{equation}
    ds^2 = -(4br|X| + 1 - 4br^2) dT^2 +
      \frac{dX^2}{4br|X| + 1 - 4br^2}
    \label{R=T-dust-outside}
  \end{equation}
at the edges $x = \pm r$, $X = \pm r (1-bt^2)$ of the fluid.
This represents a black hole if $br^2 > 1/4$ which is exactly when
the transformation \req{trans1}, \req{trans2} is not one-to-one.
This condition can also be written $\rho_0 > 1/8\pi G r^2$.
The `Schwarzschild' radius is
  \begin{equation}
    |X| = r - \frac{1}{4br}
  \end{equation}
so the dust becomes a black hole when $t = 1/2br$, which is precisely
the point at which the coordinate transformation first fails to be
one-to-one.

The system (\ref{string1},\ref{string2}) has similar solutions.
The unique vacuum solution is given by the metric \req{alpha} with
  \begin{equation}
    \alpha(x) = 1 - a e^{-Q x}
    \label{string-point-source}
  \end{equation}
and dilaton field
  \begin{equation}
    \phi = -\frac{Q}{2} x
  \end{equation}
on $\R^2$ where $a$ and $Q^2=J$ are constants of integration.
This solution has been discussed in the context of a scale-invariant
higher-derivative theory of gravity~\cite{Sch91}, $c=1$~Liouville
gravity \cite{Lug91} and a non-critical string theory \cite{MSW91,Wit91}.
This metric lacks the spatial symmetry of (\ref{point-source}) and
is of type (C) above;
indeed the curvature scalar diverges as $x\to -\infty$.
If $x$ is replaced with $|x|$ in (\ref{string-point-source}), the solution
models a point source~\cite{Ros91}.  This point source is in fact
the endpoint of the gravitational collapse of a pressureless dust.
The interior region of the collapsing solution is given by
  \begin{equation}
    ds^2 = -dt^2 +
	\left( 1-\lambda \tan \left( \frac{Q}{2}t \right) \right)^2 dx^2
    \label{string-coll-in-a}
  \end{equation}
  \begin{equation}
    \phi = \phi_0 - \ln \left(\cos \left(\frac{Q}{2}t \right) \right)
    \label{string-coll-in-b}
  \end{equation}
where $0 \leq t < \frac{2}{Q} \tan^{-1} \frac{1}{\lambda}$ and
$|x| \leq r$.  The exterior vacuum solution is
  \begin{equation}
    ds^2 = - \left( 1-ae^{-Q|X|} \right) dT^2 + \frac{dX^2}{1-ae^{-Q|X|}}
    \label{string-coll-outa}
  \end{equation}
  \begin{equation}
    \phi = -\frac{Q}{2}|X|
    \label{string-coll-outb}
  \end{equation}
where $0 \leq T$ and $|X| \leq -\frac{2}{Q}\phi_0 + \frac{1}{Q}
\ln \left[1 - (1-\xi)\tanh^2(\frac{Q}{2}T) \right]$ and $Q>0$ for
asymptotic flatness. For $\phi_0<0$, the solution (\ref{string-coll-outa},
\ref{string-coll-outb}) may be $C^0$-matched to the solution
(\ref{string-coll-in-a}, \ref{string-coll-in-b}) provided an appropriate
surface stress-energy tensor and dilaton current are included \cite{Ros91}.

Finally as an example of a metric of type (D) consider
  \begin{equation}
    ds^2 = - \cos 2 \theta dt^2 + 2 \sin 2 \theta dt dx
	    + \cos 2 \theta dx^2
    \label{angular}
  \end{equation}
on $\R^2$ where $\theta = \theta (x)$.  Note that for $\theta$
constant this is Minkowski spacetime in coordinates rotated by
the angle $\theta$.  Thus when $\theta$ varies with $x$ it
determines the tilting of the light cones throughout the spacetime.
This metric can be expressed in the form \req{alpha} by
transforming to $t' = t - \int \tan 2 \theta dx^2$, which gives
  \begin{equation}
    ds^2 = - \cos 2 \theta dt^2 + \frac{dx^2}{\cos 2 \theta} .
    \label{angular2}
  \end{equation}
For a wide variety of choices of $\theta(x)$ this metric will have no definite
sign for large $|x|$. We shall make use of this metric in the form
(\ref{angular}) below as it has no coordinate singularities.

\section{Do the Field Equations put a Restriction on the spacetime?}
\label{restrict}

In $(3+1)$-dimensional general relativity, the field equations can be
written
  \begin{equation}
    R_{ab} - \half R g_{ab} + \Lambda g_{ab} = 8 \pi G T_{ab}.
    \label{einstein}
  \end{equation}
It is clear that given any spacetime $(M,g_{ab})$ we can use
\req{einstein} to {\em define} $T_{ab}$ and then the spacetime will be a
solution to these equations for this choice of $T_{ab}$. Thus
any restrictions on the spacetime metric in general relativity are a
consequence of requiring the distribution of matter to be physically
reasonable, \ie\ to be locally causal and to respect either the weak
or dominant energy conditions. Whether or not such a property holds for
the $(1+1)$-dimensional theories of gravity considered here is the subject of
the present section.

Consider first the theory based on the equations \req{R=T} and \req{cons}.
Since the Ricci scalar couples to the trace of the stress-energy,
it is not so obvious that the same property holds.  That is, given a
spacetime $(M,g_{ab})$ does there exist a symmetric tensor field $T_{ab}$
satisfying \req{R=T} and \req{cons}?  We prove that locally such a tensor
field always exists for a sufficiently  smooth metric and give an intuitive
argument conjecturing that the global result is also true, at least for
simply connected manifolds.   Note that \req{R=T} and \req{cons} represent
three equations and that the tensor field $T_{ab}$ has three independent
components.

\begin{thm}
  Let $(M,g_{ab})$ be a $(1+1)$-dimensional spacetime and assume that
  $g_{ab}$ is $C^4$ and that $p \in M$.  Then there exists a symmetric
  tensor field $T^{ab}$ defined on a neighborhood $U\!$ of $p$ satisfying
  \req{R=T} and \req{cons}.
  \label{R=T-thm}
\end{thm}

\begin{proof}
  Since $(1+1)$-dimensional spacetimes are locally conformally flat, we
  can find a neighborhood $N$ of $p$ and a conformal factor $e^\sigma$
  where $\sigma = \sigma(t,x)$ is $C^4$, such that in $N$
  the metric can be written
  \begin{equation}
    ds^2 = e^\sigma ( -dt^2 + dx^2 ).
  \end{equation}
  In this coordinate system equation \req{R=T} becomes
  \begin{equation}
    e^{-\sigma} ( \pdb{\sigma}{x}-\pdb{\sigma}{t} ) - \Lambda =
      8 \pi G e^{\sigma} ( -T^{tt}+T^{xx} )
    \label{conf-R=T}
  \end{equation}
  and equation \req{cons} becomes
  \begin{equation}
    2 \pda{T^{tt}}{t} + 3 \pda{\sigma}{t} T^{tt} + 2 \pda{T^{tx}}{x} +
      4 \pda{\sigma}{x} T^{tx} + \pda{\sigma}{t} T^{xx} = 0
    \label{conf-cons1}
  \end{equation}
  \begin{equation}
    2 \pda{T^{xx}}{x} + 3 \pda{\sigma}{x} T^{xx} + 2 \pda{T^{tx}}{t} +
      4 \pda{\sigma}{t} T^{tx} + \pda{\sigma}{x} T^{tt} = 0.
    \label{conf-cons2}
  \end{equation}
  We now must show that this system of three equations in three
  unknowns has a solution in a neighborhood of $p$.
  Let $F=T^{tt}+T^{xx}$, $G=2T^{tx}$ and $\phi=-2\sigma$.  Then using
  \req{conf-R=T}, equations \req{conf-cons1} and \req{conf-cons2} become
  \begin{equation}
    \pda{F}{t} + \pda{G}{x} = \pda{\phi}{t} F + \pda{\phi}{x} G + M_1
    \label{conf-cons3}
  \end{equation}
  \begin{equation}
    \pda{F}{x} + \pda{G}{t} = \pda{\phi}{x} F + \pda{\phi}{t} G + M_2
    \label{conf-cons4}
  \end{equation}
  where $M_1$ and $M_2$ are expressions involving $\phi$ and its
  derivatives up to third order and $\Lambda$, and thus are $C^1$.
  Now define $\tilde{F} = F + G$ and $\tilde{G} = F - G$.  Then
  $\req{conf-cons3} + \req{conf-cons4}$ becomes
  \begin{equation}
    \pda{\tilde{F}}{t} + \pda{\tilde{F}}{x} =
      (\pda{\phi}{t} + \pda{\phi}{x}) \tilde{F} + M_1 + M_2
    \label{conf-cons5}
  \end{equation}
  and $\req{conf-cons3} - \req{conf-cons4}$ becomes
  \begin{equation}
    \pda{\tilde{G}}{t} - \pda{\tilde{G}}{x} =
      (\pda{\phi}{t} - \pda{\phi}{x}) \tilde{G} + M_1 - M_2.
    \label{conf-cons6}
  \end{equation}
  Equations \req{conf-cons3} and \req{conf-cons4} have a solution
  if and only if equations \req{conf-cons5} and \req{conf-cons6} do.
  Theorem 2-1 of \cite[Chapter 4]{Gue88} shows that \req{conf-cons5} and
  \req{conf-cons6} will have solutions in a neighborhood $U$ of $p$.
  (In fact, the theorem shows that for any $C^1$ Cauchy data defined on a
  $C^1$ non-characteristic initial curve $D$ in $N$, there exists a unique
  $C^1$ solution to \req{conf-cons5} and \req{conf-cons6} in a
  neighborhood of $D$.)
\end{proof}

For the system based on \req{string1} and \req{string2} it is clear that
\req{string1} can be used to define $T_{ab}$.
What is not so clear is whether given a spacetime $(M,g_{ab})$
there exists a scalar field $\phi$ satisfying \req{string2}.
We prove that locally such a scalar field always exists for a sufficiently
smooth metric.

\begin{thm}
  Let $(M,g_{ab})$ be a $(1+1)$-dimensional spacetime and assume that
  $g_{ab}$ is $C^2$ and that $p \in M$.  Then there exists a
  scalar field $\phi$ defined on a neighborhood $U\!$ of $p$ satisfying
  \req{string2}. \label{stringthm}
\end{thm}

\begin{proof}
  As before,
  we can find a neighborhood $N$ of $p$ and a conformal factor $e^\sigma$
  where $\sigma$ is $C^2$, such that in $N$
  the metric can be written
  \begin{equation}
    ds^2 = e^\sigma ( -dt^2 + dx^2 ).
  \end{equation}
  In this coordinate system equation \req{string2} becomes
  \begin{equation}
    -\pdb{\phi}{t} + \pdb{\phi}{x} + (\pda{\phi}{t})^2 - (\pda{\phi}{x})^2 =
      \frac{1}{4} ( R_{tt} - R_{xx} - J e^\phi ).
    \label{conf-string2a}
  \end{equation}
  Define new coordinates $t'$ and $x'$ by $t' = t + x$ and $x' = t - x$.
  Then \req{conf-string2a} can be written
  \begin{equation}
    \pdc{\phi}{x'}{t'} = \pda{\phi}{t'} \pda{\phi}{x'} +
	\frac{1}{16}(- R_{tt} + R_{xx} + J e^\phi).
    \label{conf-string2b}
  \end{equation}
  Theorem 7-1 of \cite[Chapter 4]{Gue88} shows that \req{conf-string2b}
  will have $C^2$ solutions in a neighborhood $U$ of $p$ since the
  right-hand side is a continuous function of $\pda{\phi}{t'}$,
  $\pda{\phi}{x'}$, $t'$ and $x'$ and also satisfies a Lipschitz condition
  in $\pda{\phi}{x'}$ and $\pda{\phi}{t'}$.
  Since the solution will be $C^2$, it will also satisfy \req{conf-string2a}.
  (In fact, the theorem shows that for any appropriate Cauchy data defined
  on a $C^1$ non-characteristic initial curve $D$ in $N$, there exists a
  unique solution to \req{conf-string2b} in a neighborhood of $D$.)
\end{proof}

We conjecture that the above results can be extended to any simply
connected region $S$ of a spacetime and that Theorem~\ref{R=T-thm} is
valid for $C^3$ metrics. It may be possible to construct proofs of these
conjectures by investigating the field equations
without using conformal coordinates or by using a `quilting' argument that
patches together conformal neighborhoods using the Cauchy property of the
solutions. Starting with a point $p \in S$ we find a region on which a
solution exists, by the above theorem.   Then along the boundary of this
region we apply the theorem repeatedly and use the parenthetical remarks at
the end of  Theorems~\ref{R=T-thm} and \ref{stringthm} to extend the
solution to all of the spacetime $S$.

In the rest of the paper we are not concerned with specific field
equations and deal with arbitrary spacetimes, many of which are
known solutions to the theories discussed above.  If our conjectures
are true then every spacetime we discuss below is a solution, since
each connected component of these spacetimes is simply connected.

\section{The Causal Structure of $(1+1)$-Dimensional Black Holes}

\label{bh}

Penrose diagrams are invaluable tools in the investigation of the causal
structure of a spacetime.  The causal structure of a spacetime depends
only on its conformal structure, since metrics that are related by a
conformal factor have the same light cones.  As all $(1+1)$-dimensional metrics
are conformally flat, it is straightforward to choose coordinates so
that
\begin{equation}
g=e^{-2\sigma} {\pmatrix{-1&0\cr 0&1}}
\label{eq12}
\end{equation}
for any particular timelike or spacelike region. However it is in general
not possible to carry out such a transformation for the entire two-dimensional
space, as any event horizons are located at $|\sigma| = \infty$.
In contrast to this, writing the metric in the form \req{alpha}
is especially
useful in that it clearly illustrates the full event
horizon structure of the spacetime in a manner similar to the
$(3+1)$-dimensional Schwarzchild metric. In general $\alpha(x)$ will
take on both positive and negative values, corresponding respectively to
timelike and spacelike regions. Points at which the metric changes
signature are given by $\alpha(x)=0$; these are  coordinate singularities
and locate the event horizons.

The spacetime in \req{alpha} with $\alpha=\alpha(x)$ may be maximally
extended by carrying out the Kruskal-Szekeres transformation:
\bea
uv=\exp\left[\int^x \frac{dz}{|\alpha(z)|}\right]  \label{eq14a}  \\
\frac{u}{v}= -\sgn(\alpha)\exp(t)  \label{eq14b}
\eea
which yields
\be
ds^2 =  4\alpha(uv)\frac{du dv}{uv}   \label{eq15}
\ee
where $\alpha(uv)$ is implicitly defined via eq. (\ref{eq14a}). As in the
$(3+1)$-dimensional case, the $(u,v)$ space is a double cover of the
$(t,x)$ space and the horizons are at $uv=0$.

Using the metric in the form \req{alpha} makes it easy to identify the
generic structure of curvature singularities for static
metrics. From \req{riemann} and \req{ricci} it may be shown that both
the Riemann and Ricci tensors are uniquely
determinable in terms of the Ricci scalar in two spacetime dimensions.
Since $R=-\frac{d^2\alpha}{dx^2}$,
all curvature singularities will manifest themselves in terms of
divergences in the second derivative of $\alpha(x)$. This will not
necessarily occur at a point where the metric diverges. Consider
the $(1+1)$-dimensional analog of a spherically symmetric black hole,
a black hole for which $\alpha=\alpha(|x|)$. This will correspond to
a spacetime of type (A) or (B) above.  The curvature is
\be
R = -\alpha''(r) - 2\delta(x)\alpha'(0)   \label{eq16}
\ee
where $r=|x|$ and the prime denotes $d/dr$.  If $\alpha'(0)\neq 0$ then
there will be a delta-function singularity in the curvature. If
$\alpha''(r)$ and $\alpha(0)$ are finite over the entire range of $r$
($0\leq r<\infty$) this will be the only singularity in the curvature.

Using the coordinates given in \req{eq15}, the Penrose diagram for
such a black hole may be easily constructed. The result is given in
figure~\ref{fig-penrose1}.  It is qualitatively the
  \begin{figure}
    \vspace{6.5cm}
    \caption{Penrose diagram for the $(1+1)$-dimensional spherically
    symmetric black hole}
    \label{fig-penrose1}
  \end{figure}
same as the diagram for the $(3+1)$-dimensional Schwarzchild case, except
that each point on the diagram represents a zero-sphere instead of a
2-sphere. Since a zero-sphere consists of two points, an alternative
representation of the  entire spacetime  $-\infty <x< \infty$ may be given
by taking two copies of figure~\ref{fig-penrose1}, one for $x>0$ and
the other for $x<0$, and
joining them at each of the lines at $|x|=0$ at the respective top
and bottom of each copy leading to a singular curvature at the junction.
A simple visualization is both sides of the paper.   Metrics describing
such spacetimes are given by (\ref{point-source}) with $\Lambda=0$ and
by (\ref{string-coll-outa}).

The Penrose diagram for the collapsing fluid discussed in
section~\ref{background} (\ref{R=T-dust},\ref{R=T-dust-outside})
is given in figure~\ref{fig-fluid}.
  \begin{figure}
    \vspace{6.5cm}
    \caption{Penrose diagram for a collapsing fluid}
    \label{fig-fluid}
  \end{figure}
Each point in this diagram represents a zero-sphere.  As above, this
can be represented by two copies of the figure joined along $|x|=0$.

Note that the time orientation on each copy of these diagrams is
the same, and so closed timelike curves are not present in the
spacetimes described by \ref{fig-penrose1} and \ref{fig-fluid}.
Indeed, for the collapsing fluid observers on either side of it
can travel through the fluid to verify before collapse that they
have the same time orientation; smoothness of the metric implies
that this is maintained after collapse. The static black holes
(\ref{point-source}) and (\ref{string-coll-outa}) model the
endpoint of such collapse and so are taken to have the same time
orientation for positive and negative $x$ outside the horizon.
One might consider taking multiple copies of Figure 1   and
joining them in sequence at $|x|=0$. However such joinings at
$|x|=0$  (which are not at co-ordinate singularities but are at
delta-function curvature singularities) would make the entire
manifold non-Hausdorff since each point on each individual copy
represents a zero-sphere \cite{TomRobb}.

Black hole spacetimes of type (C) have a different structure.  The Penrose
diagram is still qualitatively the same as figure~\ref{fig-penrose1},
except that each point
on the diagram represents only one point in the spacetime (as a opposed to
being a zero-sphere). Hence there is now a 1-1 mapping between points in
spacetime and points on the diagram, instead of a 2-1 mapping as before. If
no curvature singularities are present, it is possible to extend this
diagram by making multiple copies of figure~\ref{fig-penrose1} and
joining them in sequence
along the horizontal lines as in figure~\ref{fig-penrose2}.
  \begin{figure}
    \vspace{8.5cm}
    \caption{An unfolded Penrose diagram}
    \label{fig-penrose2}
  \end{figure}
Again, instead of each point
representing a zero-sphere, there would also be a 1-1 mapping between
points in the (extended) spacetime and points on the diagram.

In constructing Penrose diagrams for $(1+1)$-dimensional metrics, it is
important to note that the criterion for asymptotic flatness is slightly
more general than in higher dimensions. It is sufficient to require that
$\alpha(x)\to K|x|+C$ for large $|x|$, since a Rindler transformation may
then be applied locally to obtain a flat metric. Taking
$\alpha=\ln(\cosh(Kx))+C$, for example, satisfies this criterion;
it has no curvature singularities and its Penrose diagram is of the type
given in figure~\ref{fig-penrose2}.
In general such metrics are solutions of the field equations only for
physically unreasonable $(1+1)$-dimensional stress-energy tensors.


Diagrams for the cosmological cases may also be easily constructed.
Consider the metric (\ref{point-source}) with $M=0$.  For $C=-1$ this
yields either the metric for deSitter space if $\Lambda>0$:
\be
ds^2 = - \cos^2(\sqrt{\frac{|\Lambda|}{2}}y) dt^2 + dy^2 \label{anticos1}
\ee
(using the transformation $\sin(\sqrt{\frac{|\Lambda|}{2}}y) =
\sqrt{\frac{|\Lambda|}{2}}x$)  or anti-deSitter space if $\Lambda<0$:
\be
ds^2 = - \cosh^2(\sqrt{\frac{|\Lambda|}{2}}y) dt^2 + dy^2 \label{anticos2}
\ee
(using the transformation $\sinh(\sqrt{\frac{|\Lambda|}{2}}y) =
\sqrt{\frac{|\Lambda|}{2}}x$). The Penrose diagrams for these cases are the
same as in the $(3+1)$ dimensional case. However for $C=1$ an alternate
version of anti-deSitter space is possible with the metric
\begin{equation}
ds^2  = -(\half |\Lambda| x^2 -1) dt^2 + \frac{dx^2}{\half |\Lambda| x^2 - 1}
\label{anticos3}
  \end{equation}
which may be written as
\be
ds^2 = - \sinh^2(\sqrt{\frac{|\Lambda|}{2}}y) dt^2 + dy^2 \label{anticos4}
\ee
using the transformation $\cosh(\sqrt{\frac{|\Lambda|}{2}}y) =
\sqrt{\frac{|\Lambda|}{2}}x$.  For large $y$ the spacetime described
by (\ref{anticos4}) is the same as that described by (\ref{anticos2}).
However there is an event horizon at $y=0$ ($x = \sqrt{2/|\Lambda|}$)
which is absent in the usual anti-deSitter case.


We close this section by making some general comments on the
thermodynamics of $(1+1)$-dimensional black holes. It is
straightforward to show using either naive Wick-rotation
arguments \cite{Man90,Ros91} or a more formal quantum-field-theoretic
treatment \cite{SharDir,TomRobb} that the temperature $T$ of a black
hole is given by
\begin{equation}
  T = \frac{M}{2\pi}   \label{temphole}
\end{equation}
where $M$ is the mass-parameter.  One can then {\it define} the
entropy $S$ of the black hole via the thermodynamic relation
\cite{Man90}
\be
dM = T dS    \label{enthole}
\ee
since one can relate the mass-parameter to the energy for both of
the theories given by  (\ref{R=T},\ref{cons}) and
(\ref{string1},\ref{string2}).  In the former case one can appeal
to the Newtonian limit of the theory \cite{Man89} and in the
latter case one can compute the ADM mass \cite{Wit91}. Hence the
entropy varies logarithmically with the mass parameter
\be
S \sim \ln(\frac{M}{M_0}) \label{entropy}
\ee
where $M_0$ is a constant of integration which appears as a
fundamental mass scale in the theory; its origin presumably lies
within a fully quantized version of the $(1+1)$ dimensional
gravitation theories discussed here. A more detailed
investigation of the general thermodynamics given by
(\ref{enthole},\ref{entropy}) may be found in ref.
\cite{TomRobb}).

Relating this definition of entropy to an area parameter
associated with the black hole is somewhat more problematic.
In $(3+1)$ dimensions Hawking's area theorem says that
the area of a closed trapped surface will never decrease.  The
association of an entropy with the area of the horizon then
implies that the entropy of a black hole will never decrease in
any physical process. In
$(1+1)$ dimensions the `area' of a closed trapped surface is
meaningless because the horizon is a zero-dimensional surface.
However it may
be true that the {\em volume} of a black hole (that is, the
geodesic length enclosed by the horizon) has a similar property.
If the metric is $C^0$ within the horizon
(\ie\ if the horizon encloses only delta-function
singularities) then the geodesic length between the horizons is
well-defined and is given by
\be
\ell = \int_a^b ds       \label{geolength}
\ee
where $a$ and $b$ are any two opposing points on the horizon's worldlines.
For the static black hole
given by  (\ref{point-source}) with $\Lambda=0$ we obtain
\begin{equation}
\ell  = \int_{-\frac{1}{2M}}^{\frac{1}{2M}}
\frac{dx}{\sqrt{1-2M|x|}}  = 2
\int_{0}^{\frac{1}{2M}} \frac{dx}{\sqrt{1-2Mx}}  = \frac{2}{M}.
\end{equation}
Also, the volume of the string-theoretic black
hole  (\ref{string-coll-outa}) is given by
\begin{equation}
\ell  = \int_{-\frac{\ln 2}{2M}}^{\frac{\ln 2}{2M}}
\frac{dx}{\sqrt{2e^{2Mx}-1}}  = 2 \int_{0}^{\frac{\ln 2}{2M}}
\frac{dx}{\sqrt{2e^{2Mx}-1}}  = \frac{\pi}{2M}.
\end{equation}
Thus for these cases we find that in $(1+1)$ dimensions the
volume of a black hole {\em decreases} as matter is added, as
follows from the following dimensional arguments.
Suppose we have a static black hole in the form \req{alpha}.
Then dimensionally we must have $\alpha = \alpha(Mx)$.
Let the horizons occur at $x_L$ and $x_R$. Then we have that
\begin{equation}  \ell =\int_{x_L}^{x_R}
\frac{dx}{\sqrt{\alpha(Mx)}} = \frac{1}{M}
\int_{y_L}^{y_R} \frac{dy}{\sqrt{\alpha(y)}}
\end{equation}
where $y = Mx$.

So it seems to be a general property of static
$(1+1)$-dimensional black holes that their volume decreases as
their mass (and entropy) increases. A $(1+1)$-dimensional analog
of the area law would then involve demonstrating that in any
physical process the volume of the black hole never increases
(and hence the entropy never decreases, since it would vary
as $S \sim - \ln(V/V_0)$ \cite{Man90}). Note that for spacetimes
of type (C) there will be region of signature $(+,-)$ which is
not enclosed by two regions of signature $(-,+)$, yielding an
infinite geodesic length for such a region and a breakdown of the
entropy/length relation.  Such objects more closely resemble
cosmological event horizons than black holes, since it is
difficult to see how they could arise as the endpoint of
gravitational collapse of some distribution of matter
\cite{Ros91}.

\section{Comments on Singularity Theorems in Two Dimensions}

\label{sing}


As mentioned in the previous section, the structure of curvature
singularities is easily analyzed using the metric in the form \req{alpha}.
Extending the well-known singularity theorems \cite{Haw73,Pen65} to two
dimensions is somewhat more problematic. Consider
Penrose's 1965 theorem, which states:

\begin{thm}[Penrose 1965]
  A spacetime $(M,g_{ab})$ cannot be null geodesically complete if:
  \begin{enumerate}
    \item $R_{ab} K^a K^b \geq 0$ for all null vectors $K^a$;
      \label{e-cond}
    \item there is a non-compact Cauchy surface in $M$;
      \label{cauchy}
    \item there is a closed trapped surface in $M$.
      \label{cts}
  \end{enumerate}
\end{thm}

A proof of this theorem is given in \cite{Haw73,Pen65}.

A spacetime is said to be {\em null} (resp. {\em timelike},
{\em spacelike}) {\em geodesically complete} if all null
(resp. timelike, spacelike) geodesics can be
extended to arbitrary affine parameter values.  A spacetime is
usually said to be {\em singular} if it is not geodesically complete.

The first two conditions are easily generalized to $(1+1)$ dimensions.
Condition \ref{e-cond} is called the null energy condition
and implies in higher dimensions that the expansion of congruences of null
geodesics monotonically decreases along the geodesics.
This condition is trivially true in $(1+1)$ dimensions because
the identity $R_{ab} = \half g_{ab} R$ implies that
for all null vectors $K^a$
  \begin{equation}
    R_{ab} K^a K^b = \half g_{ab} R K^a K^b = 0.
  \end{equation}
A {\em Cauchy surface} is a spacelike hypersurface which every inextendible
non-spacelike curve intersects exactly once, a concept which way also be
extended to $(1+1)$ dimensions.  If a spacetime
admits a Cauchy surface, one can predict the state of the spacetime
at any time in the past or future if one knows the relevant data
on the surface.
See \cite{Haw73} for more discussion of these definitions.

Although the above definitions make sense for spacetimes of arbitrary
dimension ($\geq 2$, of course), the following definition only
applies to spacetimes of dimension 3 or greater.
Let $(M,g_{ab})$ be an $n$-dimensional spacetime.
A {\em closed trapped surface $S$} is a $C^2$ compact
spacelike $(n-2)$-surface without boundary such that the two families
of null geodesics orthogonal to $S$ are converging at $S$, that is,
$_1\hat\chi_{ab}g^{ab}$ and $_2\hat\chi_{ab}g^{ab}$ are negative, where
$_1\hat\chi_{ab}$ and $_2\hat\chi_{ab}$ are the two null second fundamental
forms of $S$.  Intuitively this definition is saying that the
gravitational field is so strong at $S$ that even light cannot
escape.  Extending this definition to two spacetime
dimensions is difficult in that the closed trapped surface would have to
be $0$-dimensional, most likely consisting of two distinct points.
We have been unable to find a rigorous
definition of this concept in two dimensions.  Part of the
problem is that the idea is a local one --- it only depends on
the properties of the spacetime near $S$.  But, as shown in
section~\ref{bh}, in two dimensions there exist black holes
for which spacetime is flat over large regions, so locally the event
horizon (which is a likely candidate for the closed trapped
surface) has no distinguishing properties.  Also, when a $(1+1)$-dimensional
black hole contains a singularity, the spacetime is often
disconnected and part of the event horizon is in one half and
part in the other, complicating definitions based on the `volume'
enclosed by the surface (although this may not be a serious problem for
delta-function type singularities).  And of course, the `area' of the
surface is no help since the surface is $0$-dimensional.


As an attempt to see if some form of Penrose's theorem applies in
two spacetime dimensions, we have been investigating the causal structure
of various $(1+1)$-dimensional spacetimes containing a surface
satisfying the intuitive idea behind the definition of a
closed trapped surface, and also containing a Cauchy surface.
If such a spacetime were found that was non-singular,
it would show that either
Penrose's theorem was false in two dimensions, or that the
conditions need to be strengthened.  Recall that we need not
worry about the null energy condition, as it is always true
is two dimensions.

Consider the static spacetime defined by the metric (\ref{angular}).
Although this metric is smooth for smooth choices of $\theta(x)$,
the spacetime is null geodesically incomplete whenever $\theta$ is
non-constant and $cos\,2 \theta = 0$ for some $x$.
This is seen as follows.
Let $x^a(\lambda) = (t(\lambda), x(\lambda))$ be a geodesic.
Then the geodesic equations
  \begin{equation}
    \tdb{x}{\lambda}^a +
	\Gamma^a_{bc} \tda{x}{\lambda}^b \tda{x}{\lambda}^c = 0
  \end{equation}
can be written
  \begin{equation}
    \tdb{t}{\lambda} - \tda{\theta}{x}
      \left[ \sin^2 2 \theta \, (\tda{t}{\lambda})^2
      + 2 \cos 2 \theta \, \sin 2 \theta \, \tda{t}{\lambda} \tda{x}{\lambda}
      + (\cos^2 2 \theta + 1) (\tda{x}{\lambda})^2 \right] = 0
    \label{t-geod}
  \end{equation}
  \begin{equation}
    \tdb{x}{\lambda} + \tda{\theta}{x} \sin 2 \theta
      \left[ - \cos 2 \theta \, (\tda{t}{\lambda})^2
      + 2 \sin 2 \theta \, \tda{t}{\lambda} \tda{x}{\lambda}
      + \cos^2 2 \theta \, (\tda{x}{\lambda})^2 \right] = 0.
    \label{x-geod}
  \end{equation}
Note that the factor in brackets in equation \req{x-geod} is precisely
$g_{ab} \tda{x}{\lambda}^a \tda{x}{\lambda}^b$, so for a null geodesic
this equation becomes simply $\tdb{x}{\lambda} = 0$.
Thus for non-vertical null geodesics,
we may choose $\lambda = x$ as our affine parameter.
This justifies the claim above that these spacetimes are null
geodesically incomplete when $cos 2 \theta = 0$ for some
$x$ since the vertical lines through these points are null
geodesics and so null geodesics going in the same direction
(\ie\ ``left'' or ``right'') cannot cross these lines
(by the uniqueness property of geodesics through a specific point
with a specific tangent vector).
Since $x$ is an affine parameter for these geodesics, they must be
incomplete.

Equation \req{t-geod} can be integrated for null geodesics to give
  \begin{equation}
    t = \int \tan (\theta(x) \pm \frac{\pi}{4}) \, dx + c
    \label{geod-int}
  \end{equation}
where the choice of sign selects right- or left-moving null geodesics.

An interesting choice of $\theta(x)$ is
  \begin{equation}
    \theta(x) = \tan^{-1} \frac{2x}{1+x^2}.
    \label{eq-theta}
  \end{equation}
(See figure~\ref{fig-theta}.)
  \begin{figure}
    \vspace{6.5cm}
    \caption{Plot of $\theta(x) = \tan^{-1} \frac{2x}{1+x^2}$.}
    \label{fig-theta}
  \end{figure}
This has the nice properties that it tends to $0$ as
$x \rightarrow \pm \infty$ and equals $0$ at $0$ and
$\pm \frac{\pi}{4}$ at $\pm 1$.  Thus at $\pm 1$ the
light cones are tilting inwards at 45 degrees and no
non-spacelike geodesic can leave the region $-1 \leq x \leq 1$.
Also, since the metric is independent of $t$, the distance
between the points $(t,-1)$ and $(t,1)$ is independent of $t$.
So intuitively it would seem that this is a black hole
and that for each $t$, ${(t,-1),(t,1)}$ should be called a
``closed trapped surface'', whatever that means in two dimensions.
(The independence of $t$ is important, for if we take the
portion of Minkowski spacetime with $t > 0$ in the standard
$(t,x)$ coordinates and transform to the coordinates
$(t,x')$ where $x' = x/t$ it would seem at first glance
that the resulting metric has a black hole with horizons
at $x' = \pm 1$.
But clearly the distance between $(t,x' = -1)$ and $(t,x' = 1)$
is $2t$ so the horizons do not enclose a bounded region.)

This black hole is interesting in that there are no coordinate
singularities in the metric.
(In fact, it is $C^\infty$ and non-degenerate everywhere.)
But according to the above discussion, the null geodesics are incomplete.
The integral \req{geod-int} is easy to calculate for this
choice of $\theta$ and we find that the null geodesics are
given by
  \begin{equation}
    t = \pm x + \frac{4}{1 \mp x} + 4 \ln |1 \mp x| + c.
  \end{equation}
(See figure~\ref{geods}.)
  \begin{figure}
    \vspace{6.5cm}
    \caption{
      Some right-moving null geodesics for the metric with
      $\theta(x) = \tan^{-1} \frac{2x}{1+x^2}$.
      All null geodesics can be obtained from those in the figure
      by shifting vertically and/or reflecting in the $t$-axis.
    }
    \label{geods}
  \end{figure}
Although the geodesics are unbounded as $x \rightarrow \pm 1$,
they still have finite affine length since $x$ is an affine parameter.
It is curious how different this singular behavior is from that
typically found within black holes:  the incomplete null geodesics
approach the event horizon on the opposite side of the black hole
instead of encountering a singularity at the center of the black hole,
and the metric is $C^\infty$ everywhere.  Nevertheless
Penrose's theorem is found to hold in this case and in
several other cases that we investigated.

Although we have not yet found a counterexample,
we speculate that if a reasonable definition of a closed trapped
surface in two dimensions is discovered, Penrose's theorem will be
found to be false as stated above, but will be true with a stronger
energy condition such as the weak energy condition
($R_{ab} K^a K^b \geq 0$ for all {\em non-spacelike} vectors $K^a$).
We suspect that the weak energy condition will be sufficient because
all of the examples that we experimented with that were `close' to
violating Penrose's theorem also violated the weak energy condition.

\section{Conclusions}
\label{conc}

Theories of gravitation in two spacetime dimensions possess a wealth of
solutions whose causal structure is far from trivial. Many of these have
counterparts in $(3+1)$ dimensional general relativity, but a number of
them have features which are quite distinct from the higher dimensional
case. A more complete understanding of the implications of these spacetimes
for $(1+1)$ dimensional gravity will entail a deeper exploration of the
singularities, the entropy/volume law and, ultimately, full quantization.

\section{Acknowledgments}
\label{ack}

We would like to thank Mike Morris and Marcus Kriele for many helpful
conversations and M.A.~McKiernan and D.~Siegel for advice with
section~\ref{restrict}.
Some of the integrals in this paper were evaluated using the
MAPLE symbolic algebra program developed by the Symbolic Computation
Group at the University of Waterloo.
This work was supported by the Natural Sciences and Engineering Research
Council of Canada.

\end{document}